%% file: baryon.tex
\documentclass{webofc}
\usepackage[varg]{txfonts}   
\usepackage[linesnumbered]{algorithm2e}
\usepackage{amsmath}
\usepackage{pgfplots,pgfplotstable}
\pgfplotsset{compat=1.15}
\usepgfplotslibrary{fillbetween}
\usetikzlibrary{plotmarks}

\usepackage{tabularx}

\newcommand{\bff}{\ensuremath{\mathbf f}}
\newcommand{\bg}{\ensuremath{\mathbf g}}

\newcommand{\bx}{\ensuremath{\mathbf x}}

\newcommand{\br}{\ensuremath{\mathbf r}}
\newcommand{\bj}{\ensuremath{\mathbf j}}

\newcommand{\bv}{\ensuremath{\mathbf v}}

\newcommand{\bw}{\ensuremath{\mathbf w}}
\newcommand{\bp}{\ensuremath{\mathbf p}}
\newcommand{\by}{\ensuremath{\mathbf y}}

\newcommand{\bz}{\ensuremath{\mathbf z}}

\newcommand{\bap}{\ensuremath{\vec{\mathbf{p}}}}

\DeclareMathOperator{\argmin}{argmin}

\begin{document}
\title{Efficient computation of baryon interpolating fields in Lattice QCD}
%
%

\author{\firstname{Eloy} \lastname{Romero}\inst{1}\fnsep\thanks{\email{eloy@cs.wm.edu}} \and
        \firstname{Kostas} \lastname{Orginos}\inst{2,3}\fnsep\thanks{\email{knorgi@wm.edu}}
}

\institute{Department of Computer Science, The College of William \& Mary, USA
\and
           Department of Physics, The College of William \& Mary, USA
\and
           Jefferson Laboratory, USA
          }

\abstract{
In this work we present an efficient construction of baryon interpolating fields for lattice QCD computations of two and  three point functions.
These are essential building blocks of computations of nucleon parton distribution functions (PDFs), generalized parton distribution functions (GPDs) and transverse momentum dependent distributions functions (TMDs). Lattice QCD computations of these quantities can provide additional input to assist with the global fits on experimental data for determining TMDs, GPDs and PDFs.
}
\maketitle
%


A vital component of our long term project of determining hadronic structure from lattice QCD is the ability to compute a large class of matrix elements with high precision, both statistical and systematic.
To that extent, we plan to capitalize on distillation \cite{PhysRevD.80.054506}, for constructing suitable interpolating fields for the nucleon that has already been very successful in spectroscopy computations.

The basic idea of distillation is to restrict the operators to a small subspace (the distillation basis) containing substantial contributions of the relevant eigenstates. The reduction in the rank of the operators dramatically cuts down the cost of computing all elements of the propagation matrix, allowing for measuring more complex hadron correlation functions.

Still, the amount of computation and storage overgrows with the lattice size $N$ and the rank of the distillation basis, $n$. The optimal rank of the distillation basis is determinate experimentally, but it is roughly proportional to the volume of the spatial dimensions of the lattice.  We give an idea of the costs by showing the amount of floating-point operations and the footprint memory requirements in big-O notation considering a 4D lattice that all dimension have equal size in which the volume of the spatial dimensions is proportional to the distillation basis rank, $n$. In these terms, the most expensive parts of the computations are the formation of the matrix elements, which are tensors generated from contracting matrices (basis), requiring roughly $n^{3.3}$ operations for mesons (two matrices are contracted) and $n^{4.3}$ for baryons (three matrices are contracted). The perambulators are square matrices generated by projecting the inverse of the Dirac operator and require $n^{2.3}$ operations. At the final step of the computation, matrix elements and perambulators are contracted together and that requires $n^3$ and $n^4$ operations for mesons and baryons respectively.
Table~\ref{tab:costs} details the time that each task takes for baryons in the example used along within this document.

In this project, we focus on accelerating the generation of baryon elementals, whose time dominates over the rest of the tasks. Despite the apparent simplicity of tensor contractions, developing high-performance implementations is challenging, and the efforts have to be specialized for the characteristics of the tensors and the computing device. Unlike the multiplication of matrices, few libraries, such as CFT\cite{cft} and libtensor\cite{libtensor}, provide a powerful and flexible way to specify the tensor contractions if one is willing to sacrifice some performance.

The first optimization that we propose does not reduce the number of operations, but instead reorder the operations to minimize memory requests and operands dependency, increasing the utilization of the many arithmetic-logic units available on modern CPUs. By relying on high-performance implementations of matrix-matrix multiplication, we take advantage of optimizations specific to the computing device and the dimensions of the tensors already available in these libraries.
The second approach addresses not only the computational time but also the demanding I/O requirements of the calculation. We have explored a technique that finds a sparse approximate representation of the distillation basis in a way that the baryon elementals are also sparse.

\begin{table}[t]
\caption{Asymptotic computational cost, reference time per configuration and time-slice source, and software involved in the time-consuming tasks in estimating the two point correlation functions for a lattice $32^3 \times 64$ in a single femto\cite{femto}'s node.}
\label{tab:costs}

\centering
\begin{tabular}{lccrl}
                   & Operations & Memory    & Example & \\
Computation        & cost       & footprint & time    & Main libraries \\\hline
Distillation basis & $n^{2.3}$      & $n^{2.3}$     & 0.1 h & PRIMME \\
Baryon elementals  & $n^{4.3}$      & $n^{3.3}$     &  16 h & Harom \\
Perambulators      & $100n^{2.3}$   & $100n^{2.3}$  &   3 h & Chroma, Qphix/mg\_proto \\
Contractions       & $n^{4}$        & $n^{3.3}$     & 0.1 h & Hadron, tensor
\end{tabular}
\end{table}

\subsection*{BLAS acceleration of baryon elementals}

To explain the caveats of the approach, we introduce a simplified description of the actual tensor contractions that correspond to the computation of the baryon elementals for a particular time-slice, gauge configuration, and displacements combination. The operands of the contraction include
\begin{itemize}
\item three 5D tensors (which can be the distillation basis or an operator acting on the basis), $\bv$, $\bw$, $\by$, with components on the 3D space lattice $L_N$ of dimension $N$, the three color space components $C={1,2,3}$, and the last index is the distillation basis column from 1 to $n$;
\item a 4D tensor $\bz$ (which is a phase, $z^{(\bx,l)}=e^{-i\bp\cdot \bx}$), which also has components on $L_N$ and an index for the momenta; and finally,
\item the color contraction is indicated with $\epsilon$, which is a sparse, antisymmetric tensor representing the color contraction, $\epsilon^{(1,2,3)} = 1$, and $\epsilon^{(\alpha,\alpha,\beta)} = 0$, $\epsilon^{(\alpha,\beta,\gamma)} = -\epsilon^{(\beta,\alpha,\gamma)} = -\epsilon^{(\alpha,\gamma,\beta)}$.
\end{itemize}
The tensors are operated as follows:
\begin{equation}
        B^{(i,j,k,l)} = \sum_{\bx\in L_N,\,\alpha,\beta,\gamma\in C} \epsilon^{(\alpha,\beta,\gamma)}\, v^{(\bx,\alpha,i)}\, w^{(\bx,\beta, j)}\, y^{(\bx,\gamma, k)}\, z^{(\bx,l)},\quad \text{for\ } 1\leq i,j,k\leq n,\, 1\leq l\leq M. \label{eq:baryon}
\end{equation}

We studied all alternatives for implementing the tensor contraction Eq.~\eqref{eq:baryon} by grouping the operands into two groups. Table~\ref{tab:grouping} shows all the relevant possibilities.
For instance, creating the tensor with $\epsilon$, $\bv$, $\bw$, and $\by$ and then contracted with $\bz$ requires the minimum number of floating-point operations. However, its performance is bounded by the memory bandwidth of the computing node.
In modern CPUs with many arithmetic-logic units, the grouping with better performance, despite doing three times more floating-point operations than the previous variant, contracts the auxiliary tensor $\bff$, formed with  $\epsilon$, $\bv$, and $\bw$, with the tensor $\bg$, formed with $\by$ and $\bz$.

\begin{table}
\caption{Asymptotic extra auxiliary memory and floating-point operations (FLOPs) in computing the tensor contraction at Eq.~\eqref{eq:baryon} depending on how the operands are grouped, for a lattice of volume $N^3$, and basis $\bv$, $\bw$, $\by$ of  rank $n$, and basis $\bz$ of rank $M$.}
\label{tab:grouping}
\centering
\begin{tabular}{c|rr}
Grouping                    & Aux. memory & FLOPs \\\hline
$\epsilon\ v\ w\ y\ z$      & 0                   & $9\, M\, n^3\, N^3$ \\
$(\epsilon\ v)\ (w\ y\ z)$  & $6\, M\, n^2\, N^3$ & $6\, M\, n^3\, N^3$ \\
$(\epsilon\ v\ w)\ (y\ z)$  & $3\,     n^2\, N^3$ & $3\, M\, n^3\, N^3$ \\
$(\epsilon\ v\ w\ y)\ (z)$  & $        n^3\, N^3$ & $    M\, n^3\, N^3$
\end{tabular}
\end{table}

 For contracting the tensors $\bff$ and $\bg$, we propose to bypass most of the effort of developing and maintaining a high-performance tensor contraction by relying on optimized BLAS libraries for computing matrix-matrix multiplications, such as OpenBLAS and MKL. The use of BLAS is a state-of-the-art practice in tensor contraction on CPUs\cite{DiNapoli2014} and GPUs\cite{Abdelfattah2016}.

We developed the implementation inside the library harom\footnote{\url{https://github.com/JeffersonLab/harom}}, which is part of the software suite for spectroscopy at Jefferson Laboratory.
Like the rest of the harom's code, our implementation supports shared memory (with OpenMP) and distributed memory (with MPI) paradigms.
In harom, the lattice dimensions of the operands are distributed among the processes.
So our code first contracts the local part of the basis $\bv$, $\bw$, $\by$, and $\bz$, and, in the end, a single global reduction adds the partial results at every process.
If threading is used, the threads independently work on different partitions of the $i,j$ indices of the baryon elemental.

\begin{table}
\caption{Performance comparison in time floating-point operations per seconds (GFLOPS) of the original version (Grouping $(\epsilon\ v\ w\ y)\ (z)$) and the new implementation (Grouping $(\epsilon\ v\ w)\ (y\ z)$) in computing the baryon elementals for a configuration with a lattice $32^3 \times 64$, 19 momenta, and 18 displacements on a femto's node. Reported times do not include reading/writing disk operations.
Maximum peak performance for a femto's node is 1800 GFLOPs, and bounded by memory bandwidth is 40 GFLOPS.}\label{tab:results}
\centering
\begin{tabular}{r|rr|rr}
    & \multicolumn{2}{c|}{Grouping $(\epsilon\ v\ w\ y)\ (z)$} & \multicolumn{2}{c}{Grouping $(\epsilon\ v\ w)\ (y\ z)$} \\
$n$ & Time (s) & GFLOPS & Time (s) & GFLOPS \\\hline
32  &  4,923   &   38  &      529 & 1,066 \\
64  & 33,191   &   45  &    4,048 & 1,115 \\
128 & --       &       &   29,219 & 1,235 \\
256 & --       &       &  162,542 & 1,776
\end{tabular}
\end{table}

We have tested the performance of the new implementation mostly on Intel Skylake and Phi processors. The results show that the new implementation is ten times faster on average than the original code in computing the baryon elementals.
On Tab.~\ref{tab:results}, we report the results on femto \cite{femto}, a cluster at The Collage of William \& Mary.
As the range of the distillation basis increases, the overheads of copying back and forth from the harom representation of the tensors to the formats imposed by BLAS diminishes, and the performance of the new implementation gets closer to the peak performance of the node.

A new bottleneck has appeared after the drastic reduction in computing the baryon elementals: the time for writing the baryon elementals on the global file systems starts to dominate. For a distillation basis with a $n=256$ rank, the time that takes writing the baryon elementals on disk is double that the time expended in computing them.
The approach that we introduce in the following aims at reducing not only the computational time but also the storage of the baryon elementals.

\subsection*{Blocked distillation basis}

The distillation basis consists of the eigenvectors from the lower part of the spectrum of a Laplacian-like operator $\nabla^2$,
\[
    {\nabla^2}^{(\bx,\by)} = 6\,\delta_{\bx,\by} - \sum_{\bj\in\{(1,0,0),(0,1,0),(0,0,1)\}} U_\bj^{(\bx)}\delta_{\bx+\bj,\by} + {U_\bj^{(\bx-\bj)}}^\dagger\delta_{\bx-\bj,\by},
\]
where $U$ is the gauge field restricted to a particular time-slice that may have been smeared, which is not relevant in this context.
Like the eigenvectors of the Laplacian, those eigenvectors have common components on the local scale \cite{Luscher2007}.
The use of a local support basis to approximate the lower spectrum is a critical ingredient in the construction of the prolongator and restrictor operators in multigrid \cite{brannick2008adaptive,AMGLattice}. Also, it has been used to compress the eigenvectors \cite{Clark-Jung-MGdeflation}.

The baryon elementals generated from this local-supported, sparse bases can be computed faster, and the resulting tensors are sparse also. These sparse tensors are faster to write on disk and to contract together with other tensors.

The new approach for generating the distillation basis of rank $n$ in a time-slice is as following. First, we divide the lattice in $d$ equally sized domains and restrict $f(\nabla^2)$ to each of the domains, where $f$ is function on $\Re$ whose output is non-negative real numbers. The new basis is composed by taken $n/d$ eigenvectors from each subdomain with the largest eigenvalues. The function $f$ controls the weight of the eigenvectors. For instance, if $f$ is constant, all eigenvectors matter equally. 

We avoid the evaluation of $f(\nabla^2)$ by working with a truncated approximate eigendecomposition of $f(\nabla^2)$, $f(\nabla^2)V=V\Lambda$. Then the eigendecomposition of $Vf(\nabla^2)V^\dagger$ restricted to each subdomain $s$, ${V^s}^\dagger f(\Lambda)V^s\,W^s = W^s\tilde\Lambda^s$ is related to the singular value decomposition of $V^sf(\Lambda)$:
\begin{equation}
    \left(Vf(\Lambda)V^\dagger\right)_s\bw_i^s = {\sigma_i^s}\bw_i^s \Leftrightarrow
         V^s f(\Lambda) = W^s\Sigma^s {Q^s}^\dagger.
\end{equation}

Picking more directions on each subdomain than $n/d$ increases the overlap between the resulting basis $\bw_i$ and the original basis $\bv_i$, although only $n$ directions from the over-ranked $\bw_i$ basis are going to be used as the distillation basis, $W\rho$, with $\rho^\dagger\rho=I_n$, which are the orthogonalization of the projections of $\bv_i$ onto the subspace spanned by $\bw_i$. In other words, $\rho$ is the $Q$-factor of the QR factorization of $W^\dagger V$.

\pgfplotstableread[header=false]{data/conf1000t0alpha.txt}\tablealpha

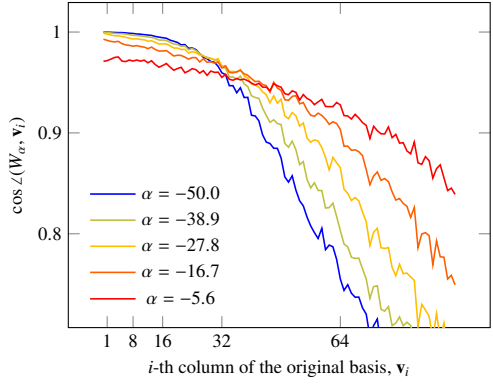
\begin{figure}[t]
\centering
\begin{scriptsize}
\begin{tikzpicture}
\begin{axis}[width=.55\textwidth, align={center}, tick label style={/pgf/number format/fixed}, 
        height=.45\textwidth,
	no marks,
        xlabel = {$i$-th column of the original basis, $\bv_i$},
        ylabel = {$\cos \angle(W_\alpha,\bv_i)$},
	ymin = 0.707,
        every axis plot/.append style={semithick},
	colormap name=hot,
	cycle list={[samples of colormap={5}]},
	xtick={1,8,16,32,64,128},
        legend entries={$\alpha=-50.0$,$\alpha=-38.9$,$\alpha=-27.8$,$\alpha=-16.7$,$\alpha=-5.6$},
        legend style={draw=none,fill=none},
	legend pos=south west
  ]
  \addplot plot table[x expr=\coordindex, y index=0] {\tablealpha};
  \addplot plot table[x expr=\coordindex, y index=1] {\tablealpha};
  \addplot plot table[x expr=\coordindex, y index=2] {\tablealpha};
  \addplot plot table[x expr=\coordindex, y index=3] {\tablealpha};
  \addplot plot table[x expr=\coordindex, y index=4] {\tablealpha};
\end{axis}
\end{tikzpicture}\end{scriptsize}

\caption{Cosine of the angles between each of the columns of the original basis, $\bv_i$, and the generated basis $W_\alpha$ with weight function $f(\lambda)=\exp(\alpha\lambda)$ for different values of $\alpha$. The spatial dimension of the lattice is $32^3$, blocking $2$ in every direction, starting with $n'=96$ and picking 128 directions ($b=16$.}
\label{fig:alpha}
\end{figure}

The function $f$ plays the role of weighing the importance of every direction.
To reinforce the presence of the smallest eigenvalues of $\nabla^2$ into the new basis, $f$ should be decreasing if we look for the largest singular values of $V^s\,f(\Lambda)$.
We have tested a simple family of functions with a single tunable parameter, $\alpha$, $f(\lambda) = \exp(\alpha\lambda)$.
Figure~\ref{fig:alpha} shows how, as $\alpha$ increases, the interior eigenvectors get more presence in detriment of the smallest eigenvectors.

When neglecting the perambulators, the hadron correlation functions are reduced to the correlation between baryon elementals at different time-slices. If $B(\bv)$ and $B'(\bv)$ are two baryon elementals at different time slices, then we want the correlations between $B(\bv)$ and $B'(\bv)$, and between $B(\bw)$ and $B'(\bw)$ to be highly correlated.
In this way, we propose the following simple heuristic to tune $\alpha$ based on the transitive property of correlations for highly correlated data: to maximize the average correlation between the baryon elementals generated with the original basis and with the new basis for a few time-slices. In principle, the correlation between the baryon elementals is not invariant under rotations of the basis, unlike the actual hadron correlation functions. We address this issue in part by selecting $\rho$ instead so that $W\rho$ is a close representation of $\bv_i$,
\[
        \rho = \argmin_{\rho^\dagger\rho=I_n} \|V - W\rho\|_F = \hat U \hat V^\dagger,
\]
where $\hat U\hat \Sigma \hat V^\dagger$ is the singular value decomposition of $V^\dagger W$.

We found that the average correlation function between baryons of both bases is concave and smooth on $\alpha$, as Fig.~\ref{fig:corr} (left) shows.
Also, the optimal value of $\alpha$ seems to depend on neither the momenta nor the displacements of the baryon elemental, and the spectra of $\nabla^2$ for different time-slices and configurations are similar enough to use the same value of $\alpha$ for all time-slices.
However, we recommend retuning $\alpha$ when changing the rank of the distillation basis as the correlation quickly drops as the rank increases, as Fig.~\ref{fig:corr} (right) shows.

\pgfplotstableread{data/corr_ranks.txt}\tablebaryon
\pgfplotstableread{data/corr_rank.txt}\tablecorrrank
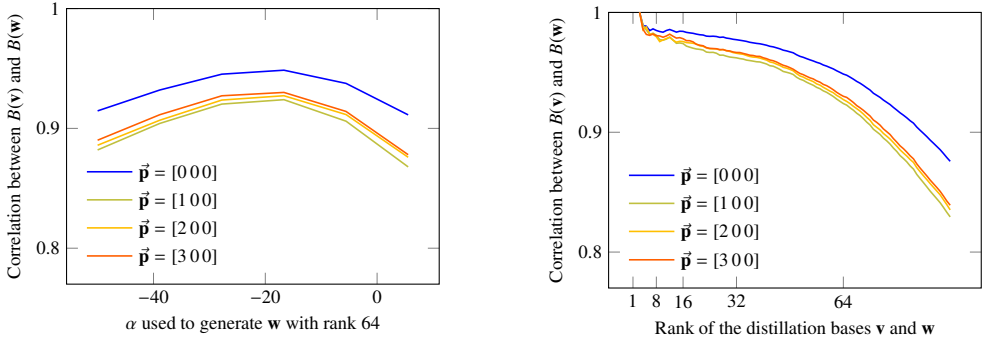
\begin{figure}
\pgfplotsset{
subgraph/.style = {
	width=.5\textwidth, align={center}, tick label style={/pgf/number format/fixed}, 
        height=.40\textwidth,
	no marks,
        ylabel = {Correlation between $B(\bv)$ and $B(\bw)$},
	ymin = 0.77,
	ymax = 1.0,
        every axis plot/.append style={semithick},
	colormap name=hot,
	cycle list={[samples of colormap={5}]},
}
}

\begin{scriptsize}
\begin{tikzpicture}
\begin{axis}[subgraph,
        xlabel = {$\alpha$ used to generate $\bw$ with rank 64},
        legend entries={{$\bap=[0\,0\,0]$},{$\bap=[1\,0\,0]$},{$\bap=[2\,0\,0]$},{$\bap=[3\,0\,0]$}},
        legend style={draw=none,fill=none},
	legend pos=south west
  ]
  \addplot plot table[x = alpha, y = mom0] {\tablecorrrank};
  \addplot plot table[x = alpha, y = mom1] {\tablecorrrank};
  \addplot plot table[x = alpha, y = mom2] {\tablecorrrank};
  \addplot plot table[x = alpha, y = mom3] {\tablecorrrank};
\end{axis}
\end{tikzpicture}\hfill
\begin{tikzpicture}
\begin{axis}[subgraph,
        xlabel = {Rank of the distillation bases $\bv$ and $\bw$},
	xtick={1,8,16,32,64,128},
        legend entries={{$\bap=[0\,0\,0]$},{$\bap=[1\,0\,0]$},{$\bap=[2\,0\,0]$},{$\bap=[3\,0\,0]$}},
        legend style={draw=none,fill=none},
	legend pos=south west
  ]
  \addplot plot table[x = index, y = mom0] {\tablebaryon};
  \addplot plot table[x = index, y = mom1] {\tablebaryon};
  \addplot plot table[x = index, y = mom2] {\tablebaryon};
  \addplot plot table[x = index, y = mom3] {\tablebaryon};
\end{axis}
\end{tikzpicture}
\end{scriptsize}

\caption{Average correlation between the baryon elementals generated with the original distillation basis $\bv$ and the blocked basis $\bw$ with weight function $f(\lambda)=\exp(\alpha\lambda)$, fixing either both bases rank to 64 (left) or the parameter $\alpha$ to the optimal value of $-16.7$ (right). The baryon elementals are of the form $B(\bv)^{(i,j,k)} = \sum_{\bx,\,\alpha,\beta,\gamma}  \epsilon^{(\alpha,\beta,\gamma)}\, v^{(\bx,\alpha,i)}\, v^{(\bx,\beta, j)}\, v^{(\bx,\gamma, k)}\, e^{-i(\bap,\bx)}$.}
\label{fig:corr}
\end{figure}

\pgfplotstableread{data/corrvv1d.txt}\tablecorr
\pgfplotstableread{data/corrvv2d_orig_3.txt}\tablecorrtwodorignear
\pgfplotstableread{data/corrvv2d_blocked_3.txt}\tablecorrtwodblockednear
\pgfplotstableread{data/corrvv2d_orig_12.txt}\tablecorrtwodorigfar
\pgfplotstableread{data/corrvv2d_blocked_12.txt}\tablecorrtwodblockedfar

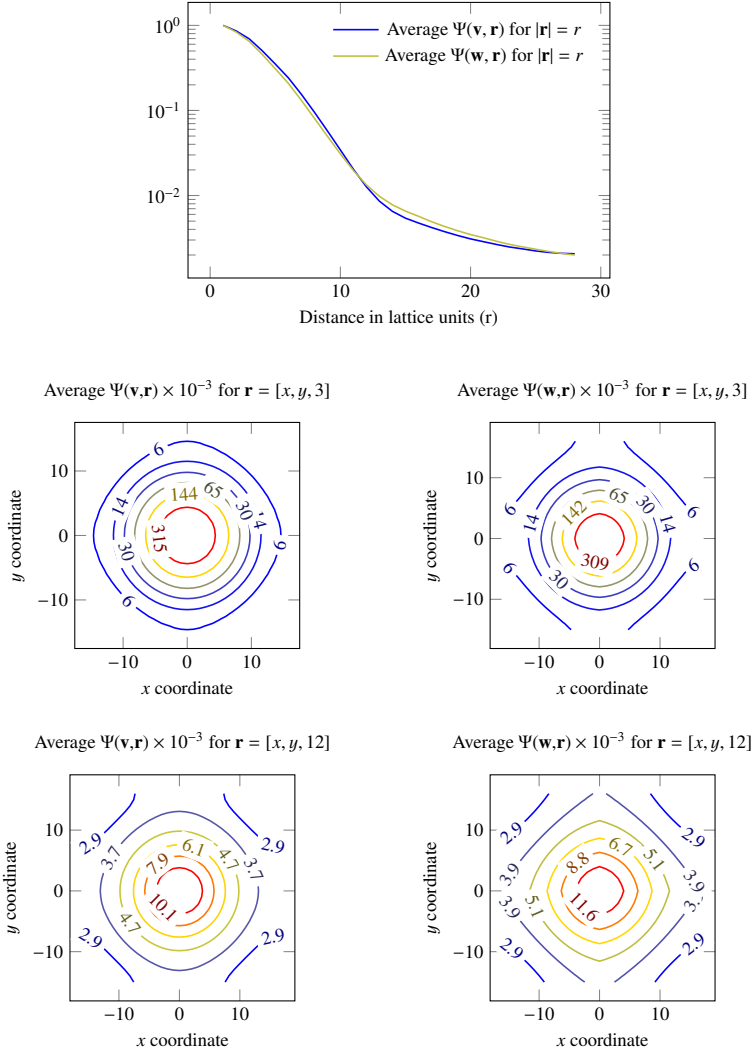
\begin{figure}
\pgfplotsset{
subgraph/.style = {
	width=.35\textwidth, align={center}, tick label style={/pgf/number format/fixed}, 
        height=.35\textwidth,
	no marks,
        xlabel = {$x$ coordinate},
        ylabel = {$y$ coordinate},
        every axis plot/.append style={semithick},
}
}

\centering
\begin{scriptsize}
\begin{tikzpicture}
\begin{semilogyaxis}[width=.55\textwidth, align={center}, tick label style={/pgf/number format/fixed}, 
        height=.4\textwidth,
	no marks,
        xlabel = {Distance in lattice units (r)},
        every axis plot/.append style={semithick},
	colormap name=hot,
	cycle list={[samples of colormap={5}]},
        legend entries={{Average $\Psi(\bv,\br)$ for $|\br|=r$},{Average $\Psi(\bw,\br)$ for $|\br|=r$}},
        legend style={draw=none,fill=none},
	legend pos=north east
  ]
  \addplot plot table[x =index, y=orig] {\tablecorr};
  \addplot plot table[x =index, y=blocked] {\tablecorr};
\end{semilogyaxis}
\end{tikzpicture}\\[5mm]
\begin{tikzpicture}
\begin{axis}[subgraph,
        title={{Average $\Psi(\bv{,}\br)\times 10^{-3}$ for $\br=[x{,}\, y{,}\, 3]$}},
  ]
  \addplot [contour prepared, contour prepared format=matlab, contour/labels=true] table {\tablecorrtwodorignear};
\end{axis}
\end{tikzpicture}\hspace*{1cm}
\begin{tikzpicture}
\begin{axis}[subgraph,
        title={{Average $\Psi(\bw{,}\br)\times 10^{-3}$ for $\br=[x{,}\, y{,}\, 3]$}},
  ]
  \addplot [contour prepared, contour prepared format=matlab, contour/labels=true] table {\tablecorrtwodblockednear};
\end{axis}
\end{tikzpicture}\\[3mm]
\begin{tikzpicture}
\begin{axis}[subgraph,
        title={{Average $\Psi(\bv{,}\br)\times 10^{-3}$ for $\br=[x{,}\, y{,}\, 12]$}},
  ]
  \addplot [contour prepared, contour prepared format=matlab, contour/labels=true] table {\tablecorrtwodorigfar};
\end{axis}
\end{tikzpicture}\hspace{1cm}
\begin{tikzpicture}
\begin{axis}[subgraph,
        title={{Average $\Psi(\bw{,}\br)\times 10^{-3}$ for $\br=[x{,}\, y{,}\, 12]$}},
  ]
  \addplot [contour prepared, contour prepared format=matlab, contour/labels=true] table {\tablecorrtwodblockedfar};
\end{axis}
\end{tikzpicture}\end{scriptsize}

\caption{Comparison of the spatial distribution of the distillation basis $\Psi$ of the original basis $\bv$ and the blocked basis $\bw$ in function of the distance (top), and at particular directions (bottom).}
\label{fig:psi}
\end{figure}

Moreover, we tested the spatial distribution of the blocked basis.
For that, we study $\Psi(\br)$ as done in \cite{PhysRevD.80.054506},
\begin{equation}
	\Psi(\bv,\br) = \sum_{\bx\in L_N} \sqrt{\sum_{\alpha\in C,\ 1\leq i\leq n}\, {\bv^{(\bx,\alpha,i)}\bv^{(\bx+\br,\alpha,i)}}^\dagger},
	\label{eq:psi}
\end{equation}
which measures the degree of smearing on the field after restricting it to the distillation basis.
At short distances, both bases seem to keep the rotational symmetry (see Fig~\ref{fig:psi} top).
At long distances, the blocked basis breaks rotational symmetry more than the original basis, as Fig.~\ref{fig:psi} bottom shows.

\input{corr_tables.tex}

\begin{figure}[t]
\pgfplotsset{
subgraph/.style = {width=.55\textwidth, align={center}, tick label style={/pgf/number format/fixed}, 
        height=.29\textwidth,
	xmin = 0, xmax = 15.3,
	ymin = .5, ymax = 2,
	cycle list={black},
},
outside fit/.style = gray!20,
inside fit/.style = gray!70,
energy/.style = {only marks,mark size=.6pt, error bars/.cd,y dir=both, y explicit},
}
\tikzset{description/.style = {anchor=west, at={(3,1.6)}, fill=white, fill opacity=0.5, text opacity=1}}

\begin{scriptsize}
\begin{tikzpicture}
\begin{axis}[subgraph]
  \addplot[name path=up , draw=none] table[x=t, y=up ] {\tablecorrpreorigzero};
  \addplot[name path=low, draw=none] table[x=t, y=low] {\tablecorrpreorigzero};
  \addplot[outside fit] fill between [of=low and up];
  \addplot[name path=up , draw=none] table[x=t, y=up ] {\tablecorrmedorigzero};
  \addplot[name path=low, draw=none] table[x=t, y=low] {\tablecorrmedorigzero};
  \addplot[inside fit] fill between [of=low and up];
  \addplot[name path=up , draw=none] table[x=t, y=up ] {\tablecorrpostorigzero};
  \addplot[name path=low, draw=none] table[x=t, y=low] {\tablecorrpostorigzero};
  \addplot[outside fit] fill between [of=low and up];
  \addplot[energy] table[x=t, y=mass, y error=err] {\tablecorrmassorigzero};
  \node [description] {$\lambda_0$ original\\$\chi^2/N_{\text{dof}}=\legendxiorigzero$, \legendenergyorigzero};
\end{axis}
\end{tikzpicture}\hfill
\begin{tikzpicture}
\begin{axis}[subgraph]
  \addplot[name path=up , draw=none] table[x=t, y=up ] {\tablecorrpreblockedzero};
  \addplot[name path=low, draw=none] table[x=t, y=low] {\tablecorrpreblockedzero};
  \addplot[outside fit] fill between [of=low and up];
  \addplot[name path=up , draw=none] table[x=t, y=up ] {\tablecorrmedblockedzero};
  \addplot[name path=low, draw=none] table[x=t, y=low] {\tablecorrmedblockedzero};
  \addplot[inside fit] fill between [of=low and up];
  \addplot[name path=up , draw=none] table[x=t, y=up ] {\tablecorrpostblockedzero};
  \addplot[name path=low, draw=none] table[x=t, y=low] {\tablecorrpostblockedzero};
  \addplot[outside fit] fill between [of=low and up];
  \addplot[energy] table[x=t, y=mass, y error=err] {\tablecorrmassblockedzero};
  \node [description] {$\lambda_0$ blocked\\$\chi^2/N_{\text{dof}}=\legendxiblockedzero$, \legendenergyblockedzero};
\end{axis}
\end{tikzpicture}

\begin{tikzpicture}
\begin{axis}[subgraph]
  \addplot[name path=up , draw=none] table[x=t, y=up ] {\tablecorrpreorigone};
  \addplot[name path=low, draw=none] table[x=t, y=low] {\tablecorrpreorigone};
  \addplot[outside fit] fill between [of=low and up];
  \addplot[name path=up , draw=none] table[x=t, y=up ] {\tablecorrmedorigone};
  \addplot[name path=low, draw=none] table[x=t, y=low] {\tablecorrmedorigone};
  \addplot[inside fit] fill between [of=low and up];
  \addplot[name path=up , draw=none] table[x=t, y=up ] {\tablecorrpostorigone};
  \addplot[name path=low, draw=none] table[x=t, y=low] {\tablecorrpostorigone};
  \addplot[outside fit] fill between [of=low and up];
  \addplot[energy] table[x=t, y=mass, y error=err] {\tablecorrmassorigone};
  \node [description] {$\lambda_1$ original\\$\chi^2/N_{\text{dof}}=\legendxiorigone$, \legendenergyorigone};
\end{axis}
\end{tikzpicture}\hfill
\begin{tikzpicture}
\begin{axis}[subgraph]
  \addplot[name path=up , draw=none] table[x=t, y=up ] {\tablecorrpreblockedone};
  \addplot[name path=low, draw=none] table[x=t, y=low] {\tablecorrpreblockedone};
  \addplot[outside fit] fill between [of=low and up];
  \addplot[name path=up , draw=none] table[x=t, y=up ] {\tablecorrmedblockedone};
  \addplot[name path=low, draw=none] table[x=t, y=low] {\tablecorrmedblockedone};
  \addplot[inside fit] fill between [of=low and up];
  \addplot[name path=up , draw=none] table[x=t, y=up ] {\tablecorrpostblockedone};
  \addplot[name path=low, draw=none] table[x=t, y=low] {\tablecorrpostblockedone};
  \addplot[outside fit] fill between [of=low and up];
  \addplot[energy] table[x=t, y=mass, y error=err] {\tablecorrmassblockedone};
  \node [description] {$\lambda_1$ blocked\\$\chi^2/N_{\text{dof}}=\legendxiblockedone$, \legendenergyblockedone};
\end{axis}
\end{tikzpicture}

\begin{tikzpicture}
\begin{axis}[subgraph]
  \addplot[name path=up , draw=none] table[x=t, y=up ] {\tablecorrpreorigtwo};
  \addplot[name path=low, draw=none] table[x=t, y=low] {\tablecorrpreorigtwo};
  \addplot[outside fit] fill between [of=low and up];
  \addplot[name path=up , draw=none] table[x=t, y=up ] {\tablecorrmedorigtwo};
  \addplot[name path=low, draw=none] table[x=t, y=low] {\tablecorrmedorigtwo};
  \addplot[inside fit] fill between [of=low and up];
  \addplot[name path=up , draw=none] table[x=t, y=up ] {\tablecorrpostorigtwo};
  \addplot[name path=low, draw=none] table[x=t, y=low] {\tablecorrpostorigtwo};
  \addplot[outside fit] fill between [of=low and up];
  \addplot[energy] table[x=t, y=mass, y error=err] {\tablecorrmassorigtwo};
  \node [description] {$\lambda_2$ original\\$\chi^2/N_{\text{dof}}=\legendxiorigtwo$, \legendenergyorigtwo};
\end{axis}
\end{tikzpicture}\hfill
\begin{tikzpicture}
\begin{axis}[subgraph]
  \addplot[name path=up , draw=none] table[x=t, y=up ] {\tablecorrpreblockedtwo};
  \addplot[name path=low, draw=none] table[x=t, y=low] {\tablecorrpreblockedtwo};
  \addplot[outside fit] fill between [of=low and up];
  \addplot[name path=up , draw=none] table[x=t, y=up ] {\tablecorrmedblockedtwo};
  \addplot[name path=low, draw=none] table[x=t, y=low] {\tablecorrmedblockedtwo};
  \addplot[inside fit] fill between [of=low and up];
  \addplot[name path=up , draw=none] table[x=t, y=up ] {\tablecorrpostblockedtwo};
  \addplot[name path=low, draw=none] table[x=t, y=low] {\tablecorrpostblockedtwo};
  \addplot[outside fit] fill between [of=low and up];
  \addplot[energy] table[x=t, y=mass, y error=err] {\tablecorrmassblockedtwo};
  \node [description] {$\lambda_2$ blocked\\$\chi^2/N_{\text{dof}}=\legendxiblockedtwo$, \legendenergyblockedtwo};
\end{axis}
\end{tikzpicture}

\begin{tikzpicture}
\begin{axis}[subgraph,xlabel={$t/a$}]
  \addplot[name path=up , draw=none] table[x=t, y=up ] {\tablecorrpreorigthree};
  \addplot[name path=low, draw=none] table[x=t, y=low] {\tablecorrpreorigthree};
  \addplot[outside fit] fill between [of=low and up];
  \addplot[name path=up , draw=none] table[x=t, y=up ] {\tablecorrmedorigthree};
  \addplot[name path=low, draw=none] table[x=t, y=low] {\tablecorrmedorigthree};
  \addplot[inside fit] fill between [of=low and up];
  \addplot[name path=up , draw=none] table[x=t, y=up ] {\tablecorrpostorigthree};
  \addplot[name path=low, draw=none] table[x=t, y=low] {\tablecorrpostorigthree};
  \addplot[outside fit] fill between [of=low and up];
  \addplot[energy] table[x=t, y=mass, y error=err] {\tablecorrmassorigthree};
  \node [description] {$\lambda_3$ original\\$\chi^2/N_{\text{dof}}=\legendxiorigthree$, \legendenergyorigthree};
\end{axis}
\end{tikzpicture}\hfill
\begin{tikzpicture}
\begin{axis}[subgraph,xlabel={$t/a$}]
  \addplot[name path=up , draw=none] table[x=t, y=up ] {\tablecorrpreblockedthree};
  \addplot[name path=low, draw=none] table[x=t, y=low] {\tablecorrpreblockedthree};
  \addplot[outside fit] fill between [of=low and up];
  \addplot[name path=up , draw=none] table[x=t, y=up ] {\tablecorrmedblockedthree};
  \addplot[name path=low, draw=none] table[x=t, y=low] {\tablecorrmedblockedthree};
  \addplot[inside fit] fill between [of=low and up];
  \addplot[name path=up , draw=none] table[x=t, y=up ] {\tablecorrpostblockedthree};
  \addplot[name path=low, draw=none] table[x=t, y=low] {\tablecorrpostblockedthree};
  \addplot[outside fit] fill between [of=low and up];
  \addplot[energy] table[x=t, y=mass, y error=err] {\tablecorrmassblockedthree};
  \node [description] {$\lambda_3$ blocked\\$\chi^2/N_{\text{dof}}=\legendxiblockedthree$, \legendenergyblockedthree};
\end{axis}
\end{tikzpicture}
\end{scriptsize}
\caption{Effective masses for time-slice shift of 5 using the original distillation basis (left) and the blocked basis (right) up to momenta 3.}
\label{fig:mass}
\end{figure}
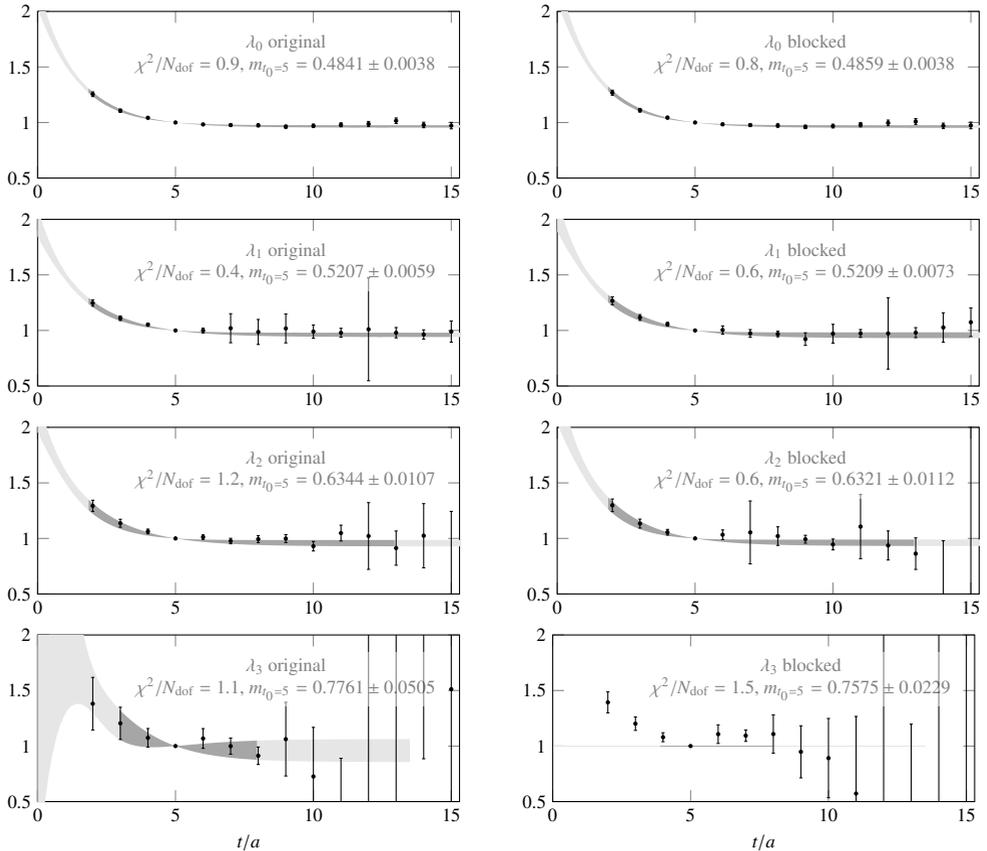

Finally, we present the effective masses using the original basis and the blocked basis in Fig.~\ref{fig:mass} up to momentum 3. Energies computed with both bases are in agreement within statistical error. Thus we do not detect systematic errors introduced by the new basis due to the more pronounced breaking of rotational symmetry. However, the new basis introduces more phisical uncertenties in estimating the mass at higher momenta.

To verify the performance benefits, we also developed a prototype code that generates the baryon elementals exploiting the sparsity patterns on the operands of the tensor contraction. The challenge in this development is controlling the overhead of manipulating the spare representation of tensors. We propose to store the nonzero blocks of the distillation basis and the tensors in a similar way as the Block Sparse Row format does for matrices (see a description at \cite{sparskit}).

It may be still useful to accelerate the contraction of tensors by the BLAS' matrix-matrix multiplication applied to the contraction of the nonzero tensor blocks. Although, some effort is required to reduce the overhead of calling the BLAS function with small matrices. This is especially dramatic for the baryon elementals with displacement, which have a more intricate nonzero pattern of small-size blocks. We got some improvement by static linking with BLAS and choosing the layout of the tensor's matrification carefully to improve data locality. But there is probably still room for improvement. We got a speedup of 2.8 in time, which is far from the expected speedup of 51 if the float-point operations limit the performance. However, the reduction in the time spent in writing the elementals on the global filesystem is 2.9, as expected. The detailed results are shown on Tab.~\ref{tab:speedups}.

\begin{table}[t]
\caption{Expected and measured speedups in the time for computing and writing on the global filesystem the baryon elementals.}
\label{tab:speedups}
\begin{center}
\begin{footnotesize}
\begin{tabular}{rl|cc|cc}
                                                                                                                                              & & \multicolumn{2}{c|}{Time speedup} & \multicolumn{2}{c}{Storage reduction} \\
\multicolumn{2}{c|}{Baryon elemental, $B(\bv)^{(i,j,k)}=\sum_{\bx,\,\alpha,\beta,\gamma}  \epsilon^{(\alpha,\beta,\gamma)}\, e^{-i(\bap,\bx)} \cdots$} & Expected & Measured & Expected & Measured \\\hline
$\cdots\, v^{(\bx,\alpha,i)}\,                         v ^{(\bx,\beta, j)}\,                                               v ^{(\bx,\gamma, k)}$ &                                                      & 64 & 5.2 & 8 & 8 \\ 
$\cdots\, v^{(\bx,\alpha,i)}\,                         v ^{(\bx,\beta, j)}\, (\mathcal{D}_{\vec\bx}                      \bv)^{(\bx,\gamma, k)}$ & for $\vec\bx\in\{\vec 1,\vec 2,\vec 3\}$             & 50 & 3.0 & 4 & 4 \\
$\cdots\, v^{(\bx,\alpha,i)}\,                         v ^{(\bx,\beta, j)}\, (\mathcal{D}_{\vec\bx} \mathcal{D}_{\vec\bx}\bv)^{(\bx,\gamma, k)}$ & for $\vec\bx\in\{\vec 1,\vec 2,\vec 3\}$             & 50 & 2.7 & 4 & 4 \\
$\cdots\, v^{(\bx,\alpha,i)}\,                         v ^{(\bx,\beta, j)}\, (\mathcal{D}_{\vec\by} \mathcal{D}_{\vec\bx}\bv)^{(\bx,\gamma, k)}$ & for $\vec\bx\neq\vec\by\in\{\vec 1,\vec 2,\vec 3\}$  & 50 & 2.6 & 2 & 2 \\
$\cdots\, v^{(\bx,\alpha,i)}\, (\mathcal{D}_{\vec\bx}\bv)^{(\bx,\beta, j)}\,                       (\mathcal{D}_{\vec\bx}\bv)^{(\bx,\gamma, k)}$ & for $\vec\bx\in\{\vec 1,\vec 2,\vec 3\}$             & 50 & 2.8 & 2 & 2 \\
$\cdots\, v^{(\bx,\alpha,i)}\, (\mathcal{D}_{\vec\by}\bv)^{(\bx,\beta, j)}\,                       (\mathcal{D}_{\vec\bx}\bv)^{(\bx,\gamma, k)}$ & for $\vec\bx<\vec\by\in\{\vec 1,\vec 2,\vec 3\}$     & 50 & 2.8 & 2 & 2 \\\hline
 & \multicolumn{1}{r|}{Average:}                                                                                                                                                                        & 51 & 2.8 & 3 & 3
\end{tabular}
\end{footnotesize}
\end{center}
\end{table}

\subsection*{Conclusions}

In this work, we propose two techniques to reduce the leading computational costs in estimating baryon correlation functions as the rank of the distillation basis increases, costs that are dominated by the generation and storage of the baryon elementals.

First, we study the performance of generating baryon elementals, which consists of the tensor contraction of the distillation basis, and we propose an implementation that maximizes the CPU utilization. Most of the fine-tuning is avoided by relying on a high-performance BLAS library.
Second, we propose to approximate the distillation basis in a spatial local support basis (blocked basis) by exploiting the local coherence of the lattice Laplacian's eigenvectors. The new basis generates very sparse baryon elementals that are more efficient to compute and store, but also contaminates the basis with higher modes that may increase the statistical uncertainty in estimating the correlation functions. Further research will evaluate the effects of the new basis at higher statistical accuracy and on other hadronic states, including excited multihadron states.

Although the asymptotic costs remain the same, the resulting reduction in costs pushes a bit further the practical limits in the lattice volume and the distillation basis' rank employed in these calculations.

\subsection*{Acknowledgements}

This work was primarily supported by the center for nuclear femtography. The gauge field configurations used in this work are from Jefferson Laboratory.
The Chroma software suite \cite{Edwards_2005} together with Qphix/mg\_proto \cite{qphix} and PRIMME \cite{PRIMME} have been also used in this work.
This work was performed in part using computing
facilities at William \& Mary which were provided by
contributions from the National Science Foundation (MRI grant
PHY-1626177), the Commonwealth of Virginia Equipment Trust Fund and
the Office of Naval Research. In particular, the majority of this work was performed on the {\em Femto} cluster at William \& Mary. 

\bibliography{bio}

\end{document}

%% file: corr_tables.tex
\pgfplotstableread{
t up low
0 2.25663 2.02756
0.205 2.07225 1.89638
0.41 1.91498 1.78039
0.615 1.781 1.67773
0.82 1.66704 1.58676
1.025 1.57025 1.50609
1.23 1.48807 1.43458
1.435 1.41816 1.37141
1.64 1.35842 1.3159
1.845 1.30706 1.26745
}\tablecorrpreorigzero
\pgfplotstableread{
t up low
1.845 1.30706 1.26745
2.05 1.26266 1.22542
2.255 1.2241 1.18913
2.46 1.19051 1.15791
2.665 1.16117 1.13111
2.87 1.13552 1.10815
3.075 1.11307 1.08851
3.28 1.09341 1.07172
3.485 1.07617 1.05739
3.69 1.06107 1.04515
3.895 1.04783 1.03472
4.1 1.03622 1.02582
4.305 1.02604 1.01825
4.51 1.01712 1.01179
4.715 1.0093 1.0063
4.92 1.00244 1.00163
5.125 0.997654 0.996429
5.33 0.994275 0.991161
5.535 0.991403 0.986543
5.74 0.988964 0.982497
5.945 0.986894 0.978951
6.15 0.985136 0.975845
6.355 0.983646 0.973123
6.56 0.982382 0.970739
6.765 0.981311 0.968652
6.97 0.980404 0.966823
7.175 0.979636 0.965222
7.38 0.978986 0.96382
7.585 0.978437 0.962593
7.79 0.977972 0.961519
7.995 0.97758 0.960578
8.2 0.977249 0.959755
8.405 0.97697 0.959035
8.61 0.976735 0.958405
8.815 0.976537 0.957854
9.02 0.97637 0.957371
9.225 0.97623 0.956949
9.43 0.976112 0.95658
9.635 0.976013 0.956257
9.84 0.97593 0.955975
10.045 0.975861 0.955728
10.25 0.975803 0.955513
10.455 0.975755 0.955324
10.66 0.975714 0.955159
10.865 0.975681 0.955015
11.07 0.975653 0.954889
11.275 0.975629 0.954779
11.48 0.97561 0.954683
11.685 0.975594 0.954598
11.89 0.975581 0.954525
12.095 0.97557 0.954461
12.3 0.975561 0.954405
12.505 0.975554 0.954356
12.71 0.975548 0.954313
12.915 0.975543 0.954276
13.12 0.975539 0.954243
13.325 0.975536 0.954214
13.53 0.975534 0.954189
13.735 0.975532 0.954168
13.94 0.97553 0.954149
14.145 0.975529 0.954132
14.35 0.975528 0.954118
14.555 0.975527 0.954105
14.76 0.975527 0.954094
14.965 0.975526 0.954084
}\tablecorrmedorigzero
\pgfplotstableread{
t up low
14.965 0.975526 0.954084
15.17 0.975526 0.954076
15.375 0.975526 0.954069
15.58 0.975526 0.954062
15.785 0.975526 0.954057
15.99 0.975526 0.954052
16.195 0.975526 0.954048
16.4 0.975526 0.954044
16.605 0.975526 0.954041
16.81 0.975526 0.954038
17.015 0.975526 0.954035
17.22 0.975526 0.954033
17.425 0.975526 0.954031
17.63 0.975526 0.95403
17.835 0.975526 0.954028
18.04 0.975526 0.954027
18.245 0.975526 0.954026
18.45 0.975526 0.954025
18.655 0.975527 0.954024
18.86 0.975527 0.954023
19.065 0.975527 0.954023
19.27 0.975527 0.954022
19.475 0.975527 0.954022
19.68 0.975527 0.954021
19.885 0.975527 0.954021
20.09 0.975527 0.954021
20.295 0.975527 0.95402
20.5 0.975527 0.95402
}\tablecorrpostorigzero
\pgfplotstableread{
t mass err
2 1.2547 0.0207142
3 1.10594 0.014608
4 1.04271 0.00947171
6 0.982343 0.0090898
7 0.97692 0.0106035
8 0.974277 0.013033
9 0.962183 0.0136935
10 0.970241 0.0146749
11 0.979186 0.017562
12 0.986852 0.0224335
13 1.01554 0.0250937
14 0.977768 0.0247174
15 0.971869 0.0290076
5 1 2.80291e-15
}\tablecorrmassorigzero
\newcommand{\legendxiorigzero}{0.9}
\newcommand{\legendenergyorigzero}{$m_{t_0=5}=0.4841\pm 0.0038$}
\pgfplotstableread{
t up low
0 2.09905 1.84616
0.205 1.94199 1.75103
0.41 1.8078 1.66446
0.615 1.69357 1.58544
0.82 1.59681 1.51303
1.025 1.51522 1.44653
1.23 1.44641 1.3857
1.435 1.38789 1.33072
1.64 1.33747 1.28179
1.845 1.29348 1.23881
}\tablecorrpreorigone
\pgfplotstableread{
t up low
1.845 1.29348 1.23881
2.05 1.25475 1.20141
2.255 1.22046 1.16904
2.46 1.19 1.14116
2.665 1.16289 1.11719
2.87 1.13872 1.09664
3.075 1.11717 1.07905
3.28 1.09794 1.06401
3.485 1.08079 1.05116
3.69 1.06548 1.04021
3.895 1.05183 1.03087
4.1 1.03965 1.02293
4.305 1.02878 1.01617
4.51 1.0191 1.01043
4.715 1.01046 1.00556
4.92 1.00277 1.00143
5.125 0.99794 0.995914
5.33 0.994987 0.989808
5.535 0.992493 0.984369
5.74 0.990392 0.979528
5.945 0.988622 0.975218
6.15 0.987136 0.971383
6.355 0.985888 0.967972
6.56 0.984844 0.964937
6.765 0.983972 0.962238
6.97 0.983245 0.959839
7.175 0.982641 0.957707
7.38 0.98214 0.955812
7.585 0.981726 0.954128
7.79 0.981386 0.952633
7.995 0.981106 0.951305
8.2 0.980878 0.950125
8.405 0.980693 0.949078
8.61 0.980544 0.948149
8.815 0.980424 0.947324
9.02 0.980329 0.946592
9.225 0.980255 0.945943
9.43 0.980197 0.945367
9.635 0.980153 0.944856
9.84 0.98012 0.944404
10.045 0.980096 0.944002
10.25 0.98008 0.943646
10.455 0.980069 0.943331
10.66 0.980063 0.943051
10.865 0.980061 0.942804
11.07 0.980062 0.942584
11.275 0.980065 0.94239
11.48 0.980069 0.942218
11.685 0.980075 0.942065
11.89 0.980081 0.94193
12.095 0.980088 0.941811
12.3 0.980096 0.941705
12.505 0.980103 0.941611
12.71 0.98011 0.941528
12.915 0.980118 0.941455
13.12 0.980125 0.94139
13.325 0.980132 0.941332
13.53 0.980138 0.941282
13.735 0.980144 0.941237
13.94 0.98015 0.941197
14.145 0.980155 0.941162
14.35 0.98016 0.941131
14.555 0.980165 0.941103
14.76 0.980169 0.941079
14.965 0.980173 0.941057
}\tablecorrmedorigone
\pgfplotstableread{
t up low
14.965 0.980173 0.941057
15.17 0.980177 0.941038
15.375 0.98018 0.941022
15.58 0.980183 0.941007
15.785 0.980186 0.940994
15.99 0.980189 0.940982
16.195 0.980191 0.940972
16.4 0.980193 0.940963
16.605 0.980195 0.940955
16.81 0.980197 0.940948
17.015 0.980199 0.940941
17.22 0.9802 0.940936
17.425 0.980201 0.940931
17.63 0.980203 0.940927
17.835 0.980204 0.940923
18.04 0.980205 0.94092
18.245 0.980206 0.940917
18.45 0.980206 0.940914
18.655 0.980207 0.940912
18.86 0.980208 0.94091
19.065 0.980208 0.940908
19.27 0.980209 0.940906
19.475 0.980209 0.940905
19.68 0.98021 0.940904
19.885 0.98021 0.940903
20.09 0.98021 0.940902
20.295 0.980211 0.940901
20.5 0.980211 0.9409
}\tablecorrpostorigone
\pgfplotstableread{
t mass err
2 1.24704 0.0282664
3 1.10998 0.0206177
4 1.05276 0.0132666
6 0.999491 0.0218477
7 1.02006 0.130563
8 0.986466 0.112475
9 1.01809 0.130032
10 0.989864 0.0592321
11 0.979107 0.0402171
13 0.979802 0.0465563
14 0.963954 0.0416994
15 0.989909 0.0953271
5 1 1.40662e-14
12 1.01034 0.462907
}\tablecorrmassorigone
\newcommand{\legendxiorigone}{0.4}
\newcommand{\legendenergyorigone}{$m_{t_0=5}=0.5207\pm 0.0059$}
\pgfplotstableread{
t up low
0 2.28933 1.95749
0.185 2.11745 1.86064
0.37 1.97015 1.77018
0.555 1.84457 1.68541
0.74 1.73803 1.60583
0.925 1.64775 1.53144
1.11 1.57079 1.46271
1.295 1.50437 1.40023
1.48 1.44625 1.34429
1.665 1.39482 1.29481
1.85 1.34894 1.25142
}\tablecorrpreorigtwo
\pgfplotstableread{
t up low
1.85 1.34894 1.25142
2.035 1.30779 1.21358
2.22 1.27075 1.18072
2.405 1.23735 1.15228
2.59 1.2072 1.12772
2.775 1.17995 1.10656
2.96 1.15532 1.08835
3.145 1.13306 1.07272
3.33 1.11293 1.05931
3.515 1.09474 1.04783
3.7 1.0783 1.03802
3.885 1.06345 1.02964
4.07 1.05004 1.0225
4.255 1.03793 1.01643
4.44 1.02699 1.01127
4.625 1.01713 1.0069
4.81 1.00823 1.0032
4.995 1.00021 1.00008
5.18 0.99745 0.992971
5.365 0.995243 0.986451
5.55 0.993397 0.980576
5.735 0.991856 0.975285
5.92 0.990575 0.97052
6.105 0.989515 0.966231
6.29 0.988641 0.962371
6.475 0.987926 0.958898
6.66 0.987344 0.955774
6.845 0.986874 0.952965
7.03 0.986498 0.95044
7.215 0.9862 0.94817
7.4 0.985968 0.946131
7.585 0.98579 0.944298
7.77 0.985657 0.942652
7.955 0.985561 0.941174
8.14 0.985496 0.939847
8.325 0.985454 0.938656
8.51 0.985433 0.937587
8.695 0.985427 0.936628
8.88 0.985434 0.935767
9.065 0.98545 0.934995
9.25 0.985474 0.934303
9.435 0.985503 0.933683
9.62 0.985535 0.933126
9.805 0.985571 0.932627
9.99 0.985608 0.93218
10.175 0.985646 0.93178
10.36 0.985683 0.931421
10.545 0.985721 0.9311
10.73 0.985757 0.930812
10.915 0.985793 0.930554
11.1 0.985827 0.930323
11.285 0.985859 0.930116
11.47 0.98589 0.929931
11.655 0.985919 0.929766
11.84 0.985947 0.929618
12.025 0.985973 0.929485
12.21 0.985997 0.929366
12.395 0.986019 0.92926
12.58 0.98604 0.929165
12.765 0.98606 0.92908
12.95 0.986078 0.929004
}\tablecorrmedorigtwo
\pgfplotstableread{
t up low
12.95 0.986078 0.929004
13.135 0.986094 0.928936
13.32 0.986109 0.928876
13.505 0.986124 0.928821
13.69 0.986137 0.928773
13.875 0.986149 0.928729
14.06 0.986159 0.928691
14.245 0.986169 0.928656
14.43 0.986179 0.928625
14.615 0.986187 0.928597
14.8 0.986195 0.928573
14.985 0.986202 0.928551
15.17 0.986208 0.928531
15.355 0.986214 0.928513
15.54 0.986219 0.928497
15.725 0.986224 0.928483
15.91 0.986228 0.928471
16.095 0.986232 0.92846
16.28 0.986236 0.92845
16.465 0.986239 0.928441
16.65 0.986242 0.928433
16.835 0.986245 0.928426
17.02 0.986247 0.928419
17.205 0.986249 0.928414
17.39 0.986251 0.928409
17.575 0.986253 0.928404
17.76 0.986255 0.9284
17.945 0.986256 0.928396
18.13 0.986257 0.928393
18.315 0.986259 0.92839
18.5 0.98626 0.928388
}\tablecorrpostorigtwo
\pgfplotstableread{
t mass err
2 1.29326 0.0505691
3 1.13678 0.0356959
4 1.0627 0.0211394
6 1.01161 0.0202921
7 0.976335 0.0245906
8 0.993598 0.0313117
9 0.99842 0.0360002
10 0.929878 0.0428668
11 1.04811 0.0711825
13 0.912947 0.153993
5 1 9.97522e-15
12 1.02177 0.300381
14 1.02439 0.288871
15 0.345501 0.89774
}\tablecorrmassorigtwo
\newcommand{\legendxiorigtwo}{1.2}
\newcommand{\legendenergyorigtwo}{$m_{t_0=5}=0.6344\pm 0.0107$}
\pgfplotstableread{
t up low
0 3 0
0.135 3 0.211145
0.27 3 0.505714
0.405 3 0.741677
0.54 3 0.928275
0.675 3 1.07338
0.81 3 1.18367
0.945 3 1.26478
1.08 3 1.32145
1.215 2.7035 1.3576
1.35 2.44749 1.37645
1.485 2.23248 1.38063
1.62 2.05342 1.37224
1.755 1.9059 1.35311
1.89 1.78575 1.32508
2.025 1.68871 1.29036
2.16 1.61024 1.25169
2.295 1.54579 1.21197
2.43 1.49141 1.17371
2.565 1.44412 1.13863
2.7 1.40186 1.10763
2.835 1.36333 1.08101
2.97 1.32771 1.05868
}\tablecorrpreorigthree
\pgfplotstableread{
t up low
2.97 1.32771 1.05868
3.105 1.29451 1.04034
3.24 1.26342 1.02559
3.375 1.23425 1.01399
3.51 1.20684 1.00513
3.645 1.18111 0.998592
3.78 1.15697 0.994028
3.915 1.13436 0.99111
4.05 1.1132 0.989554
4.185 1.09343 0.989113
4.32 1.07499 0.989573
4.455 1.05781 0.990753
4.59 1.04183 0.992496
4.725 1.02698 0.994671
4.86 1.01321 0.997169
4.995 1.00045 0.999896
5.13 1.00278 0.988646
5.265 1.00575 0.97773
5.4 1.00875 0.967649
5.535 1.01176 0.958349
5.67 1.01473 0.949777
5.805 1.01764 0.941882
5.94 1.02046 0.934618
6.075 1.02319 0.927938
6.21 1.02581 0.921801
6.345 1.02831 0.916165
6.48 1.0307 0.910995
6.615 1.03296 0.906253
6.75 1.03509 0.901907
6.885 1.0371 0.897926
7.02 1.039 0.894281
7.155 1.04077 0.890946
7.29 1.04243 0.887895
7.425 1.04398 0.885106
7.56 1.04543 0.882557
7.695 1.04677 0.880228
7.83 1.04803 0.878101
7.965 1.04919 0.87616
}\tablecorrmedorigthree
\pgfplotstableread{
t up low
7.965 1.04919 0.87616
8.1 1.05027 0.874389
8.235 1.05126 0.872772
8.37 1.05219 0.871298
8.505 1.05304 0.869955
8.64 1.05383 0.86873
8.775 1.05456 0.867613
8.91 1.05523 0.866596
9.045 1.05584 0.86567
9.18 1.05641 0.864827
9.315 1.05693 0.864059
9.45 1.05741 0.86336
9.585 1.05785 0.862724
9.72 1.05826 0.862145
9.855 1.05863 0.861619
9.99 1.05897 0.86114
10.125 1.05928 0.860704
10.26 1.05957 0.860308
10.395 1.05983 0.859948
10.53 1.06007 0.859621
10.665 1.06029 0.859323
10.8 1.06049 0.859053
10.935 1.06068 0.858807
11.07 1.06084 0.858584
11.205 1.061 0.858381
11.34 1.06114 0.858197
11.475 1.06127 0.858029
11.61 1.06138 0.857877
11.745 1.06149 0.857739
11.88 1.06159 0.857614
12.015 1.06168 0.8575
12.15 1.06176 0.857397
12.285 1.06183 0.857303
12.42 1.0619 0.857218
12.555 1.06196 0.85714
12.69 1.06202 0.85707
12.825 1.06207 0.857006
12.96 1.06212 0.856948
13.095 1.06216 0.856896
13.23 1.0622 0.856848
13.365 1.06223 0.856805
13.5 1.06227 0.856766
}\tablecorrpostorigthree
\pgfplotstableread{
t mass err
3 1.20441 0.144919
4 1.07399 0.0836052
6 1.06828 0.0885278
7 0.99941 0.072421
8 0.912589 0.0785621
2 1.38103 0.236844
5 1 1.35883e-14
9 1.06214 0.331639
10 0.726011 0.442238
11 0.280955 0.6091
12 2.36955 4.02825
13 5.20523 7.30655
14 4.42493 3.53958
15 1.51133 16.276
}\tablecorrmassorigthree
\newcommand{\legendxiorigthree}{1.1}
\newcommand{\legendenergyorigthree}{$m_{t_0=5}=0.7761\pm 0.0505$}
\pgfplotstableread{
t up low
0 2.35819 2.08884
0.205 2.15604 1.95006
0.41 1.98399 1.82731
0.615 1.83774 1.71864
0.82 1.71364 1.62232
1.025 1.60854 1.53682
1.23 1.51963 1.46093
1.435 1.44433 1.39373
1.64 1.38027 1.33456
1.845 1.32541 1.28286
}\tablecorrpreblockedzero
\pgfplotstableread{
t up low
1.845 1.32541 1.28286
2.05 1.27813 1.23801
2.255 1.23715 1.19933
2.46 1.20149 1.16609
2.665 1.17039 1.13763
2.87 1.14321 1.11329
3.075 1.11944 1.09253
3.28 1.09863 1.07482
3.485 1.0804 1.05975
3.69 1.06444 1.04692
3.895 1.05045 1.036
4.1 1.03819 1.02673
4.305 1.02745 1.01885
4.51 1.01804 1.01216
4.715 1.00979 1.00649
4.92 1.00257 1.00167
5.125 0.997592 0.996242
5.33 0.994134 0.990699
5.535 0.991205 0.985844
5.74 0.988726 0.981593
5.945 0.986628 0.977871
6.15 0.984855 0.974612
6.355 0.983356 0.97176
6.56 0.98209 0.969263
6.765 0.981021 0.967078
6.97 0.980119 0.965166
7.175 0.979359 0.963493
7.38 0.978719 0.962029
7.585 0.978179 0.960749
7.79 0.977726 0.95963
7.995 0.977345 0.95865
8.2 0.977024 0.957794
8.405 0.976756 0.957046
8.61 0.97653 0.956392
8.815 0.976342 0.95582
9.02 0.976184 0.95532
9.225 0.976052 0.954883
9.43 0.975942 0.954501
9.635 0.97585 0.954168
9.84 0.975774 0.953876
10.045 0.97571 0.953622
10.25 0.975657 0.9534
10.455 0.975614 0.953205
10.66 0.975578 0.953036
10.865 0.975548 0.952888
11.07 0.975523 0.952759
11.275 0.975503 0.952646
11.48 0.975487 0.952547
11.685 0.975473 0.952461
11.89 0.975462 0.952386
12.095 0.975454 0.952321
12.3 0.975446 0.952264
12.505 0.975441 0.952214
12.71 0.975436 0.95217
12.915 0.975432 0.952132
13.12 0.97543 0.952099
13.325 0.975427 0.95207
13.53 0.975426 0.952045
13.735 0.975424 0.952023
13.94 0.975424 0.952004
14.145 0.975423 0.951987
14.35 0.975422 0.951973
14.555 0.975422 0.95196
14.76 0.975422 0.951949
14.965 0.975422 0.951939
}\tablecorrmedblockedzero
\pgfplotstableread{
t up low
14.965 0.975422 0.951939
15.17 0.975422 0.951931
15.375 0.975422 0.951923
15.58 0.975422 0.951917
15.785 0.975422 0.951912
15.99 0.975422 0.951907
16.195 0.975423 0.951902
16.4 0.975423 0.951899
16.605 0.975423 0.951896
16.81 0.975423 0.951893
17.015 0.975423 0.95189
17.22 0.975423 0.951888
17.425 0.975424 0.951886
17.63 0.975424 0.951885
17.835 0.975424 0.951883
18.04 0.975424 0.951882
18.245 0.975424 0.951881
18.45 0.975424 0.95188
18.655 0.975424 0.951879
18.86 0.975424 0.951879
19.065 0.975424 0.951878
19.27 0.975425 0.951877
19.475 0.975425 0.951877
19.68 0.975425 0.951876
19.885 0.975425 0.951876
20.09 0.975425 0.951876
20.295 0.975425 0.951876
20.5 0.975425 0.951875
}\tablecorrpostblockedzero
\pgfplotstableread{
t mass err
2 1.26835 0.0226598
3 1.11025 0.0163714
4 1.04389 0.0111828
6 0.983257 0.0103947
7 0.976761 0.0119053
8 0.97242 0.0147173
9 0.960436 0.0154272
10 0.967723 0.0166386
11 0.980223 0.0197188
12 0.996637 0.0250025
13 1.00836 0.0262062
14 0.969993 0.024428
15 0.973532 0.0301044
5 1 3.18279e-15
}\tablecorrmassblockedzero
\newcommand{\legendxiblockedzero}{0.8}
\newcommand{\legendenergyblockedzero}{$m_{t_0=5}=0.4859\pm 0.0038$}
\pgfplotstableread{
t up low
0 2.19684 1.89181
0.205 2.02186 1.79679
0.41 1.87296 1.70923
0.615 1.74697 1.62811
0.82 1.64137 1.55233
1.025 1.55377 1.48107
1.23 1.48119 1.41452
1.435 1.42004 1.35381
1.64 1.36731 1.29982
1.845 1.32102 1.25264
}\tablecorrpreblockedone
\pgfplotstableread{
t up low
1.845 1.32102 1.25264
2.05 1.27995 1.21184
2.255 1.24332 1.17678
2.46 1.21054 1.14676
2.665 1.18117 1.12113
2.87 1.15483 1.0993
3.075 1.1312 1.08073
3.28 1.11001 1.06497
3.485 1.091 1.05161
3.69 1.07396 1.04031
3.895 1.05869 1.03075
4.1 1.045 1.02269
4.305 1.03275 1.0159
4.51 1.02177 1.01019
4.715 1.01195 1.0054
4.92 1.00317 1.00138
5.125 0.998027 0.995316
5.33 0.995229 0.988295
5.535 0.992903 0.982021
5.74 0.990975 0.976417
5.945 0.989381 0.971414
6.15 0.988069 0.966947
6.355 0.986993 0.962962
6.56 0.986114 0.959407
6.765 0.985401 0.956237
6.97 0.984826 0.95341
7.175 0.984366 0.950892
7.38 0.984001 0.948647
7.585 0.983714 0.946648
7.79 0.983492 0.944868
7.995 0.983323 0.943283
8.2 0.983198 0.941873
8.405 0.983107 0.940618
8.61 0.983045 0.939501
8.815 0.983006 0.938508
9.02 0.982985 0.937625
9.225 0.982978 0.936839
9.43 0.982982 0.936142
9.635 0.982995 0.935521
9.84 0.983014 0.93497
10.045 0.983037 0.934481
10.25 0.983063 0.934046
10.455 0.983091 0.93366
10.66 0.983121 0.933317
10.865 0.98315 0.933013
11.07 0.98318 0.932743
11.275 0.983209 0.932503
11.48 0.983237 0.932291
11.685 0.983264 0.932102
11.89 0.98329 0.931935
12.095 0.983314 0.931786
12.3 0.983337 0.931655
12.505 0.983359 0.931538
12.71 0.983379 0.931435
12.915 0.983397 0.931343
13.12 0.983415 0.931262
13.325 0.98343 0.93119
13.53 0.983445 0.931126
13.735 0.983459 0.931069
13.94 0.983471 0.931019
14.145 0.983482 0.930975
14.35 0.983493 0.930936
14.555 0.983502 0.930901
14.76 0.983511 0.93087
14.965 0.983519 0.930843
}\tablecorrmedblockedone
\pgfplotstableread{
t up low
14.965 0.983519 0.930843
15.17 0.983526 0.930819
15.375 0.983532 0.930797
15.58 0.983538 0.930779
15.785 0.983543 0.930762
15.99 0.983548 0.930747
16.195 0.983552 0.930734
16.4 0.983556 0.930722
16.605 0.98356 0.930712
16.81 0.983563 0.930703
17.015 0.983566 0.930695
17.22 0.983568 0.930688
17.425 0.983571 0.930682
17.63 0.983573 0.930676
17.835 0.983575 0.930671
18.04 0.983576 0.930667
18.245 0.983578 0.930663
18.45 0.983579 0.93066
18.655 0.98358 0.930657
18.86 0.983581 0.930654
19.065 0.983582 0.930652
19.27 0.983583 0.93065
19.475 0.983584 0.930648
19.68 0.983585 0.930646
19.885 0.983585 0.930645
20.09 0.983586 0.930643
20.295 0.983586 0.930642
20.5 0.983587 0.930641
}\tablecorrpostblockedone
\pgfplotstableread{
t mass err
2 1.26683 0.0359344
3 1.11657 0.0263369
4 1.0566 0.0164002
6 1.00385 0.03734
7 0.973679 0.0340677
8 0.968347 0.0249714
9 0.922181 0.0562748
10 0.970896 0.0843145
11 0.973177 0.0350434
13 0.980631 0.0449772
14 1.02708 0.131978
15 1.07392 0.12876
5 1 9.04368e-15
12 0.973554 0.321299
}\tablecorrmassblockedone
\newcommand{\legendxiblockedone}{0.6}
\newcommand{\legendenergyblockedone}{$m_{t_0=5}=0.5209\pm 0.0073$}
\pgfplotstableread{
t up low
0 2.48654 2.03063
0.185 2.27614 1.92537
0.37 2.09708 1.82718
0.555 1.94541 1.73532
0.74 1.81767 1.64915
0.925 1.71054 1.56841
1.11 1.62053 1.49339
1.295 1.54417 1.42479
1.48 1.47843 1.3632
1.665 1.42103 1.30877
1.85 1.37035 1.26121
}\tablecorrpreblockedtwo
\pgfplotstableread{
t up low
1.85 1.37035 1.26121
2.035 1.32528 1.21999
2.22 1.28501 1.18447
2.405 1.24891 1.15399
2.59 1.21649 1.12793
2.775 1.18736 1.10569
2.96 1.16116 1.08677
3.145 1.13759 1.0707
3.33 1.11639 1.05709
3.515 1.09733 1.04557
3.7 1.08018 1.03585
3.885 1.06477 1.02767
4.07 1.05092 1.02079
4.255 1.03848 1.01502
4.44 1.02731 1.0102
4.625 1.01728 1.00618
4.81 1.00828 1.00284
4.995 1.00021 1.00007
5.18 0.997783 0.992967
5.365 0.995906 0.986479
5.55 0.994371 0.980665
5.735 0.993123 0.975456
5.92 0.992116 0.970793
6.105 0.991309 0.966618
6.29 0.990668 0.962882
6.475 0.990166 0.95954
6.66 0.989778 0.956551
6.845 0.989484 0.953879
7.03 0.989266 0.95149
7.215 0.989111 0.949356
7.4 0.989007 0.947449
7.585 0.988943 0.945746
7.77 0.988911 0.944225
7.955 0.988905 0.942867
8.14 0.988919 0.941656
8.325 0.988948 0.940575
8.51 0.988988 0.939611
8.695 0.989036 0.93875
8.88 0.989089 0.937983
9.065 0.989146 0.9373
9.25 0.989205 0.93669
9.435 0.989264 0.936147
9.62 0.989323 0.935663
9.805 0.989381 0.935232
9.99 0.989437 0.934848
10.175 0.989491 0.934506
10.36 0.989542 0.934202
10.545 0.989591 0.93393
10.73 0.989637 0.933689
10.915 0.98968 0.933474
11.1 0.98972 0.933283
11.285 0.989757 0.933113
11.47 0.989792 0.932962
11.655 0.989824 0.932828
11.84 0.989854 0.932708
12.025 0.989882 0.932602
12.21 0.989907 0.932507
12.395 0.98993 0.932423
12.58 0.989951 0.932348
12.765 0.98997 0.932282
12.95 0.989988 0.932223
}\tablecorrmedblockedtwo
\pgfplotstableread{
t up low
12.95 0.989988 0.932223
13.135 0.990004 0.93217
13.32 0.990019 0.932124
13.505 0.990032 0.932082
13.69 0.990045 0.932046
13.875 0.990056 0.932013
14.06 0.990066 0.931984
14.245 0.990075 0.931958
14.43 0.990083 0.931935
14.615 0.99009 0.931915
14.8 0.990097 0.931897
14.985 0.990103 0.931881
15.17 0.990108 0.931867
15.355 0.990113 0.931854
15.54 0.990118 0.931843
15.725 0.990122 0.931833
15.91 0.990125 0.931824
16.095 0.990129 0.931816
16.28 0.990132 0.93181
16.465 0.990134 0.931803
16.65 0.990136 0.931798
16.835 0.990139 0.931793
17.02 0.99014 0.931789
17.205 0.990142 0.931785
17.39 0.990144 0.931782
17.575 0.990145 0.931779
17.76 0.990146 0.931776
17.945 0.990147 0.931774
18.13 0.990148 0.931772
18.315 0.990149 0.93177
18.5 0.99015 0.931768
}\tablecorrpostblockedtwo
\pgfplotstableread{
t mass err
2 1.29827 0.0568219
3 1.13377 0.0401415
4 1.05558 0.0256717
6 1.03336 0.0433114
8 1.02098 0.0838787
9 0.992906 0.0351945
10 0.945721 0.04892
12 0.936488 0.130823
13 0.862952 0.142063
5 1 1.71063e-14
7 1.05406 0.283037
11 1.1064 0.288943
14 0.483897 0.495109
15 0.119313 2.05996
}\tablecorrmassblockedtwo
\newcommand{\legendxiblockedtwo}{0.6}
\newcommand{\legendenergyblockedtwo}{$m_{t_0=5}=0.6321\pm 0.0112$}
\pgfplotstableread{
t up low
0 1 1
0.135 1 1
0.27 1 1
0.405 1 1
0.54 1 1
0.675 1 1
0.81 1 1
0.945 1 1
1.08 1 1
1.215 1 1
1.35 1 1
1.485 1 1
1.62 1 1
1.755 1 1
1.89 1 1
2.025 1 1
2.16 1 1
2.295 1 1
2.43 1 1
2.565 1 1
2.7 1 1
2.835 1 1
2.97 1 1
3.105 1 1
3.24 1 1
3.375 1 1
3.51 1 1
3.645 1 1
3.78 1 1
3.915 1 1
}\tablecorrpreblockedthree
\pgfplotstableread{
t up low
3.915 1 1
4.05 1 1
4.185 1 1
4.32 1 1
4.455 1 1
4.59 1 1
4.725 1 1
4.86 1 1
4.995 1 1
5.13 1 1
5.265 1 1
5.4 1 1
5.535 1 1
5.67 1 1
5.805 1 1
5.94 1 1
6.075 1 1
6.21 1 1
6.345 1 1
6.48 1 1
6.615 1 1
6.75 1 1
6.885 1 1
7.02 1 1
7.155 1 1
7.29 1 1
7.425 1 1
7.56 1 1
7.695 1 1
7.83 1 1
7.965 1 1
}\tablecorrmedblockedthree
\pgfplotstableread{
t up low
7.965 1 1
8.1 1 1
8.235 1 1
8.37 1 1
8.505 1 1
8.64 1 1
8.775 1 1
8.91 1 1
9.045 1 1
9.18 1 1
9.315 1 1
9.45 1 1
9.585 1 1
9.72 1 1
9.855 1 1
9.99 1 1
10.125 1 1
10.26 1 1
10.395 1 1
10.53 1 1
10.665 1 1
10.8 1 1
10.935 1 1
11.07 1 1
11.205 1 1
11.34 1 1
11.475 1 1
11.61 1 1
11.745 1 1
11.88 1 1
12.015 1 1
12.15 1 1
12.285 1 1
12.42 1 1
12.555 1 1
12.69 1 1
12.825 1 1
12.96 1 1
13.095 1 1
13.23 1 1
13.365 1 1
13.5 1 1
}\tablecorrpostblockedthree
\pgfplotstableread{
t mass err
4 1.07938 0.040115
6 1.1071 0.0836836
7 1.09367 0.0513225
8 1.10812 0.171975
2 1.3932 0.0944377
3 1.20123 0.0600491
5 1 1.21792e-14
9 0.948095 0.234363
10 0.891637 0.356615
11 0.572711 0.694014
12 2.05396 1.61127
13 -0.208835 1.40583
14 2.17758 2.04503
15 -5.0533 18.4728
}\tablecorrmassblockedthree
\newcommand{\legendxiblockedthree}{1.5}
\newcommand{\legendenergyblockedthree}{$m_{t_0=5}=0.7575\pm 0.0229$}